\magnification=\magstep1
\baselineskip=14pt
\overfullrule=0 pt
\hsize = 15 true cm
\vsize = 22 true cm
\font\titulobold=cmbx12 scaled\magstep1

\vskip 1 true cm

\titulobold

\titulobold
\centerline{A General Approach to the Modelling} 
\centerline{of Trophic Chains} 
\tenrm
\bigskip
\centerline{\bf Rui Dil\~ao$^a$\footnote\dag{\rm rui@sd.ist.utl.pt}\ \ {\tenrm and}\ \ Tiago Domingos$^b$\footnote\ddag{\rm fdomin@alfa.ist.utl.pt}}
\bigskip
\centerline{\it $^a$Grupo de Din\^amica N\~ao-Linear, Department  of Physics}
\centerline{\it $^b$Grupo de Din\^amica N\~ao-Linear, Department  of Mechanical Engineering}
\centerline{\it Instituto Superior T\'ecnico, Av. Rovisco Pais,1049-001 Lisboa Codex, Portugal}

\bigskip
\bigskip

\centerline{\bf Abstract}

Based on the law of mass action (and its microscopic foundation) and mass conservation,  we present here a method to derive consistent dynamic models for the time evolution of systems with an arbitrary number of species.
 Equations are derived through a mechanistic description, ensuring that all parameters have ecological meaning.
After discussing the biological mechanisms associated 
to the  logistic and Lotka-Volterra equations, we show how to derive
general
models for trophic chains, including the effects of  internal states at fast
time scales.
We show that conformity with the mass action law leads to different
functional forms for the Lotka-Volterra and trophic chain models. We use
mass conservation  to recover the concept of carrying capacity for
an arbitrary food chain.

\bigskip

\vfill
\bigskip

\vfill


\noindent {\bf Fax:} (351)-1-8419123.
\bigskip

\eject
{\bf 1. Introduction}
\medskip

There exists a multitude of models for trophic interactions. For example, Royama (1971) and May (1974) describe different alternatives to model the same  interactions, and Berryman {\it et al.} (1995a) give a table of twenty five 
alternatives to model predator-prey systems. However, the calibration and validation of these models with experimental and observational data is systematicaly lacking, and most ecologists prefer to adjust time series data with empirical models that have no connection to the specific ecological processes (Solow, 1995). 

For single-species population dynamics, the logistic equation is the basic paradigm,  introduced in almost any ecology textbook.  It accurately  predicts population densities in systems such as bacterial batch cultures (Schlegel, 1992) and human populations (Banks, 1994) and, when generalized,  describes the dynamics of many single species populations in both laboratory and field (Gause, 1934; Allee {\it et al.}, 1949; Thomas {\it et al.}, 1980; Berryman and Millstein, 1990). It is applicable to multiple situations in ecology and biology (Banks, 1994) and bioeconomics (Clark, 1990). However,
the logistic equation  has  been criticized with the argument that the underlying 
carrying capacity concept has no
mechanistic meaning, being simply a fitting parameter (Kooi {\it et al.}, 1998), and obscuring the relation between population growth and resource availability (Getz, 1984).  

For multi-species population dynamics, the basic model is the Lotka-Volterra equations. They are the basis of almost all the theory of trophic interactions. A further development of these equations was the recognition that there exist limits to the capacity for consumption, leading to the introduction of the Holling functional response, commonly called Type II (Holling, 1959). It was later verified that the functional form of this curve coincides with the Monod function used in microbiology and the Michaelis-Menten mechanism in enzyme kinetics. The Holling Type II functional response lies at the heart of current trophic chain dynamics theory (Oksanen {\it et al.}, 1981).

In 1928 Volterra adopted the mass action principle of chemical kinetics
 to write the dynamic equations for the densities of a prey-predator
system (Berryman, 1992).  In 1977, Nicolis and Prigogine  showed that the logistic equation could be derived in analogy with  chemical kinetics, using the mass action law and a mass conservation principle. The mass action law lies at the heart of most population dynamics theory, as in epidemiology (Anderson and May, 1991) and in structured population models (Metz and Diekmann, 1986; Metz and de Roos, 1992).

Mass conservation is a controversial issue in population dynamics. Some authors have argued that population dynamics models do not have to conform to mass conservation (e.g. Berryman {\it et al.}, 1995b). However, the effect of mass conservation on the dynamics of communities has been ascertained (De Angelis 
{\it et al.}, 1989).

On the other hand, one of the problems with the complexity of ecological systems is that there may exist internal states of the systems which we cannot measure
(Arditi and Ginzburg, 1989). There is a systematic way of eliminating these variables from the system description, if their dynamics occur at  faster time scales than the time scale of population dynamics. In physics this is called the adiabatic approximation (Haken, 1983), and in chemistry it is called the quasi-steady state assumption (Segel, 1988; Segel and Slemrod, 1989; Borghans {\it et al.}, 1996; Stiefenhofer, 1998). This approach has been used in ecology  to distinguish between different time scales (O'Neill {\it et al.}, 1986; Michalski {\it et al.}, 1997).

In this paper, we take the  chemical kinetics analogy to its full consequences, showing how to derive  the population dynamic equations of an arbitrary food web from the ecological mechanisms of interaction. This approach is based on the fact that both organisms and molecules are discrete entities that interact with each other. The advantage of this analogy is that we can bring from physics to ecology  the knowledge of statistical mechanics about the transition from individual motions to macroscopic behavior, deriving the precise limits of validity of the deterministic population dynamics description (Maurer, 1998). This program unifies the deterministic with the stochastic approach to population dynamics, as far as local densities of individuals do not fluctuate too much around the average density of the whole population.

This paper is organized as follows. In the next section we review the main 
techniques for deriving evolution equations of chemical kinetics, with the  necessary modifications for ecological systems. The general evolution equations are taken in accordance to the mass action law. We then impose a  conservation law which is equivalent to the assumption of closedness of the ecological system. 
This mass conservation law makes it possible to model a renewable resource, leading  to the concept of carrying capacity.
To account for internal  states occurring at fast  time scales, we introduce  
the mechanism that leads to a Michaelis-Menten resource uptake,
which can be compared with  logistic type mechanisms.
In section 5, we show that the correct application of the mass action law
leads to a new Lotka-Volterra type system of equations, for two-species and $n$-species interactions.
In section 6, we derive the general form of the evolution equations for  
a trophic chain where species can have internal states.  
Taking limits of ecologically
significant parameters, more complicated models 
are reduced exactly to simpler ones.

\bigskip

{\bf 2. From chemical kinetics to ecological mechanisms: the mass action law}
\medskip

In order to construct population dynamics models with 
clear and  ecologically significant mechanisms and precise conditions 
of validity and applicability,  we now introduce the analytical  
technique that will be used in this paper. 

At the scale of interatomic distances, the motion of molecules in a solution 
is random. When two  eventually binding chemical species collide, a new molecule appears, decreasing the mole number of the initial chemical species and increasing the mole number of the newly formed chemical species. Analogously,
in population dynamics, if we assume randomness in the motion of individuals,
the interaction of individuals with a resource is a collision.
At collision, the individual can consume the resource --- binding --- or
simply  ignore it.  
Therefore,  both systems can be considered similar and, at the macroscopic level, the
mean densities (mole number per unit volume, in chemical kinetics; number of individuals per
unit  area or volume, in population dynamics) are  described by
the same evolution laws.

So, we consider a closed
area (or volume) $S$ with several species or resources, $A_j$, $j=1,\ldots ,m$, with number of individuals given by $n_j$. Interactions in  $S$  are described by the collision diagrams
$$
\nu_{i1} A_1+\cdots +\nu_{im} A_m\to^{r_i} 
\mu_{i1} A_1+\cdots +\mu_{im} A_m\, ,\quad  i=1,\ldots ,n \eqno(2.1)
$$
where $\nu_{ij}$ and $\mu_{ij}$ are positive parameters measuring the number
of individuals that are consumed and produced, with $\nu_{ij}$ being integers, and
the constants $r_i$  measure the  rate of the interaction. 

Suppose further that the species $A_j$ is well distributed in $S$ with 
mean density $a_j=n_j/S$. It follows from a master equation approach (van Kampen,
1992, pp. 166-172) that
$$
{da_j\over dt}=\sum_{i=1}^{n} r_i(\mu_{ij}-\nu_{ij}) 
a_1^{\nu_{i1}}\ldots a_m^{\nu_{im}}
\, ,\quad  j=1,\ldots ,m\eqno(2.2)
$$

Equation (2.2) expresses the mass action law and is derived using the following assumptions,  establishing its limits of validity. 
  
\item{i)} At each instant of time, the densities of each species are approximately constant 
over the finite territory $S$.
\item{ii)} The densities are low.
\item{iii)} Individual motions are independent of each other,
in such a way that the  collision frequency is proportional
to the product of the probability densities  of finding the different individuals in a small region.
\item{iv)} Interaction probabilities are independent of the past history of 
organisms.
\item{v)} The motion of individuals is random ans is due to some form of 
collision with the environment, as in Brownian motion  (Haken, 1983).

Adopting this analogy between ecological and chemical interactions, it 
is possible to derive the usual interaction laws found in the ecological
literature, with the advantage that now these evolution laws have 
a precise mechanistic meaning given by the collision diagrams (2.1).

In the following, we will make an  additional simplification.
Defining the order of a reaction as $\sum_{i=1}^n \nu_{ij}$, we will only consider second or lower order reactions, i.e., $\sum_{i=1}^n \nu_{ij}\leq 2$, since higher order reactions  have comparatively negligible probabilities of occurrence.

In the following and to simplify the notation, we will represent  
species  and species densities in diagram (2.1) and equations (2.2) by
the same symbol $A_j$.

\bigskip

{\bf 3. Logistic   autotrophs}
\medskip

Let us represent by $N$ the density of individuals of a species per unit of area or volume.
Let us represent  resources by $A$. Schematically we can represent 
the ecological interaction --- species consuming resources and reproducing --- by the following diagram
$$
A+N\to^{r_0} (1+e) N
\eqno(3.1)
$$
where $e>0$ is a constant expressing the increase in species density, and  $r_0$ is
a rate constant expressing the velocity of the transformation, at the population 
dynamics time scale. 

Based on the mass action law of \S 2, the time evolution associated to the transformation (3.1) is  
$$
\eqalign{
{d N\over dt}&= r_0eAN \cr
{d A\over dt}&= -r_0AN \cr
}
\eqno(3.2) 
$$
Multiplying by $e$ the second equation in (3.2) and adding to the first one,
it follows that the time variation 
of $N(t)+eA(t)$ is zero, and, therefore, $N(t)+eA(t)=constant$.
With $K=N(0)+eA(0)$, and eliminating $A$ from equations (3.2), we obtain
$$
{d N\over dt} = r_0N(K-N):=rN\left(1-{N\over K}\right) \eqno(3.3)  
$$
where $K$ is the carrying capacity, and $r=r_0K$ is the intrinsic
growth rate of the population. 
The species dynamics (3.3) has the solution $N(t)=KN(0)e^{rt}/(K+(e^{rt}-1)N(0))$, and predictions about the  
values of the density of a population can be obtained by fitting a time series 
with the explicit solution $N(t)$. In the limit $t\to \infty$, $N(t)\to K$.

In order to make the mechanism (3.1) more realistic, we introduce death rate
occurring at the ecological time scale: 
$$\eqalign{
A+N&\to^{r_0} (1+e) N\cr
N&\to^{d}  \beta A \cr}
\eqno(3.4)
$$
where the second diagram represents the death of individual with death rate $d$,
and $\beta $ is a recycling constant determined below. The time evolution of the transformation (3.4) is  now
$$
\eqalign{
{d N\over dt}&= r_0eAN-dN \cr
{d A\over dt}&= -r_0AN+d\beta N \cr
}
\eqno(3.5) 
$$
Imposing a conservation law of the form, $\gamma A+N=constant$, and introducing (3.5) into the equation
$\gamma \dot A+\dot N=0$, we obtain, $\gamma =e$
and $\beta =1/e$. Therefore, with $K=N(0)+e A(0)$, and eliminating $A$ from equations (3.5), we obtain
$$
{d N\over dt}= r_0N(K-N)-dN:=rN\left(1-{N\over K}\right)-dN 
\eqno(3.6) 
$$
where $K$ is the carrying capacity, $r=r_0K$ is the intrinsic growth rate
of the population, and $d$ is the death rate. 
If $r>d$, (3.6) has a stable equilibrium solution for $N=K(r-d)/r$. If $r<d$, the only (stable) nonnegative solution is $N=0$. But now, the carrying capacity parameter 
$K$ is not the equilibrium value attained by the population in the limit $t\to \infty$, instead it is the value of the conservation law associated to (3.5).
In this case, the solution
of (3.6) is $N(t)=e^{rt}K(d-r)/(Ce^{dt}-re^{rt})$, where $C=r+K(d-r)/N(0)$, 
and time series fitting of observational data is straightforward.

The 
recycling condition $N\to \beta A$ has been introduced in (3.4) 
in order to have a conservation law,
which leads to the decoupling of system (3.6), and the determination of an explicit
solution. But, for example, in bacterial batch cultures, where the logistic
equation is tested, the recycling condition is not verified. 
In these systems, after the bacterial population has exhausted the resources,
the population density decreases, a feature that can not be obtained with the recycling condition in the
logistic equation.

Dropping the recycling condition in (3.4), we obtain the mechanism
$$
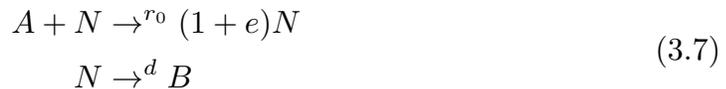
\eqalign{
A+N&\to^{r_0} (1+e) N\cr
N&\to^{d} B\cr}
\eqno(3.7)
$$
where $B$ is some nonrecycling resource. In this case, the evolution equations
associated to $A$ and $N$ are
$$
\eqalign{
{d N\over dt}&= r_0eAN-dN \cr
{d A\over dt}&= -r_0AN \cr}
\eqno(3.8) 
$$
but no conservation law exists enabling its integration.
Therefore, the fitting parameters are more difficult to estimate.
Clearly, the system $\{A,N  \}$ is open, but the system $\{A,N,B\}$ is
closed and has a conservation law ($e A+N+B=constant$).

In Fig. 1 we show 
the graph of the solution $N(t)$  of the logistic type models
(3.3), (3.6) and (3.8).
Comparing the qualitative behavior of the three systems, we take the following
conclusions.
\item{i)} For large values of available resources $A$, the solutions
of the three systems are quantitatively similar.
\item{ii)} The carrying capacity parameter $K$ only coincides with
the equilibrium value of the population if there are no deaths and
populations remain constant after exhausting the resources.
\item{iii)} In the exponential growth phase, 
the three models give qualitatively and quantitatively similar 
results. 
\item{iv)} If the death rate is small compared with the rate 
constant $r_0$, the maximum density of populations calculated 
by the three logistic models is approximated
by the carrying capacity $K$, and the population densities in the exponential phase of growth are similar.

In the following, we will always keep   the biomass recycling 
hypothesis, leading to a conservation law and, therefore, to a carrying capacity. 

Specifying intermediate internal states in the life cycle of a species,
we now show that, under a steady state approximation, we also obtain 
a logistic equation. 

Suppose that the species $N$  has $n$ behavioral internal states, $N_1,\ldots 
,N_n$, and  that reproduction occurs according to the mechanism
$$\eqalign{
A+N_1 \mathrel{\mathop \rightarrow^{r_1}}   N_2 \qquad &N_1 \mathrel{\mathop \rightarrow^{d}} \beta_1A \cr
A+N_2 \mathrel{\mathop \rightarrow^{r_2}}  N_3 \qquad &N_2 \mathrel{\mathop \rightarrow^{d}} \beta_2A \cr
\vdots &\cr
A+N_n \mathrel{\mathop \rightarrow^{r_n}} (1+e) N_1 \qquad &N_n \mathrel{\mathop \rightarrow^{d}} \beta_nA \cr
}\eqno(3.9)
$$
where the $\beta_i $ are constants to be determined later, in order to introduce a conservation law. The dynamical equations of mechanism (3.9) are
$$
\eqalign{
{d A\over dt}&= -\sum_{i=1}^{n}r_i A N_i +d\sum_{i=1}^{n}\beta_i N_i\cr
{d N_1\over dt}&= -r_1 AN_1 +r_n (1+e)AN_n-d N_1\cr
{d N_i\over dt}&= r_{i-1} AN_{i-1} -r_iA N_i-d N_i\, ,\quad  i=2,\ldots ,n\cr
}\eqno(3.10)
$$

We now impose the conservation law,
$$
{d A\over dt}+ \gamma_1 {d N_1\over dt}+\cdots +\gamma_n {d N_n\over dt}=0\eqno(3.11)
$$
Introducing (3.10) into (3.11), and solving this equation for any 
$A$ and $N_i$, we obtain the 
parameter values
$$
\gamma_i={n\over e}+i-1\, , \, \beta_i=\gamma_i\, ,\, i=1,\ldots ,n\eqno(3.12)
$$
and, from (3.11), the conservation law is 
$$
A+\gamma_1N_1+\ldots +\gamma_nN_n=K\eqno(3.13)
$$

We now introduce a steady state assumption, over the internal states,
$$
{dN_i\over dt}=0\, , i=2, \ldots ,n\eqno(3.14)
$$
Solving equations (3.14), we obtain,
$$
N_i={r_{i-1}A\over r_{i}A+d}N_{i-1}={r_{i-1}\over r_{i}+d/A}N_{i-1}\, , i=2, \ldots ,n\eqno(3.15)
$$
where, by (3.13), $A$ is a function of the $N_i'$. But, taking the limit, $d\to 0$, we obtain, $N_i=r_{i-1}N_{i-1}/r_i$, which, by induction, gives,
$$
N_i={r_1\over r_i}N_1\, , i=2, \ldots ,n\eqno(3.16)
$$
and, with $N=N_1+\ldots + N_n$, by (3.10), (3.11), (3.13) and (3.16),
$$
{dN\over dt}=er_nAN_n-dN={e\over {1\over r_1}+\cdots +{1\over r_n}}N\left(K-N\sum_{i=1}^{n}
{\gamma_i\over r_i({1\over r_1}+\cdots +{1\over r_n}})\right) -dN
\eqno(3.17)
$$
which is the logistic equation.

Therefore, at the ecological scale, internal intermediate states do not
introduce further dynamical changes in population dynamics 
equations, as far as the death rate $d$ is small. 

For general systems with several basic nutrients, one conservation
law for each resource should be introduced.

\bigskip

{\bf 4. Monod autotrophs}
\medskip

In order to derive the mechanisms for Monod type population dynamics models, we consider a system with an autotroph $N$, which can be
found in two states: searching for nutrient, $N_{s}$, and processing (handling) nutrient $N_{h}$, with $N=N_{s}+N_{h}$. When the autotrophs find nutrient, they switch from searching to  handling, increasing their biomass. At a behavioral time scale, handling autotrophs decay to searching autotrophs, and reproduce. The death rate is the same  for
both handling and searching autotrophs.  The kinetic mechanism is thus:
$$\eqalign{
A+N_{s} \mathrel{\mathop \rightarrow^{r_1}}  (1+e)N_{h} \qquad &N_{h} \mathrel{\mathop \rightarrow^{d}} \beta A \cr
N_{h} \mathrel{\mathop \rightarrow^{r_2}}  N_{s} \qquad &N_{s} \mathrel{\mathop \rightarrow^{d}} \beta A \cr
}\eqno(4.1)
$$
where $e>0$ is the conversion constant  accounting for the increase
in species density, $r_1$ and $r_2$ are the  rates at which
processes occur, and $\beta$ is a recycling constant that will be
calculated later under the assumptions of a  conservation law.
The resource density is  $A$.
 
Let us now apply the formalism of \S 2 to the interaction given by diagram (4.1). By (2.1) and (2.2), we obtain
$$
\eqalign{
{d A\over dt}&= -r_1 AN_s +d\beta (N_s+N_h)\cr
{d N_s\over dt}&= -r_1 AN_s +r_2 N_h-d N_s\cr
{d N_h\over dt}&= r_1 (1+e)AN_s -r_2 N_h-d N_h\cr
}\eqno(4.2)
$$
as evolution equations of resources and individuals.

From the point of view of population dynamics, we
can count species numbers but resources are difficult to 
estimate. Therefore, in order to apply and compare the predictions
of system (4.2) with a real system, we must be able
to rewrite (4.2) without  modelling explicitly the resource density $A$.
As  in the logistic equation,
the only way to do this is to impose a conservation law, 
$$
\gamma {d A\over dt}+ {d N_s\over dt}+{d N_h\over dt}=0\eqno(4.3)
$$
Introducing (4.3) into (4.2), and solving for the parameters, 
we obtain,
$$
\gamma=e\, , \, 
\beta= {1\over e}\eqno(4.4)
$$
and the conservation law, $eA+N_s+N_h=constant=K$.
Under these conditions, the system described by the mechanism (4.1) with
$\beta$ given by (4.4) is,
$$
\eqalign{
{d N_s\over dt}&= -{r_1\over e} N_s (K- N_s-N_h)+r_2 N_h-d N_s\cr
{d N_h\over dt}&= {r_1 (1+e)\over e} N_s (K- N_s-N_h)-r_2 N_h-d N_h\cr
}\eqno(4.5)
$$
where $K=e A+ N_s+N_h$ is the carrying capacity. 

But, at the ecological time scale, our goal is to follow the time evolution 
of the total density of a population, $N=N_s+N_h$, without knowledge
of behavioral states of the population. Therefore, with $N=N_s+N_h$, 
system (4.5) is rewritten as,
$$
\eqalign{
{d N\over dt}&= r_1  (N- N_h)(K -N) -d N\cr
{d N_h\over dt}&= {r_1 (1+e)\over e} (N-N_h) (K- N)-r_2 N_h-d N_h\cr
}\eqno(4.6)
$$

It is natural to assume that there are two time scales in this problem: at the ecological time scale the dynamics of $N_h$ is so fast that 
$N_h$ is  constant.  This allows us to apply the steady state assumption,
$$
{dN_{h} \over dt}=0\eqno(4.7)
$$
implying that, at the ecological scale, 
$$ 
{dN \over dt}= r_1N\left(K-{N}\right) {1 \over 
1+{r_1(1+e)\over e(d+r_2)}(K-N)}-d N :=r_1N(K -N){1\over 1+\delta (K -N)}-dN\eqno(4.8)
$$
where $\delta=r_1(1+e)/ e(d+r_2)$. 
System (4.8) has an equilibrium
state for $N=(Kr_1-d (1+\delta K))/(r_1-\delta d)$. This equilibrium state equals
the value of the carrying capacity $K$ when $d\to 0$.

If we take the behavioral dynamics in (4.1) infinitely fast when compared with the ecological
time scale, $r_2\to \infty$, $\delta\to 0$ and (4.8)  reduces 
to the logistic equation (3.6).
 For the steady state assumption (4.7) to be valid, the dynamics of $N_{h}$ must be much faster than the dynamics of $N$, which is simply
obtained for large $r_2$. 

In Fig. 2, we compare 
density growth curves of the one species  model (4.8) with the 
logistic model (3.6), for the same parameters values. The effect
of the introduction of an intermediate state,  delays the growth and slightly changes the steady state. 
This contrasts with model (3.9), where the introduction of intermediate behavioral states corresponding to the different states of the life cycle 
of an individual do not change the overall time behavior of the
population at the ecological time scale.

\bigskip
{\bf 5. Modified Lotka-Volterra trophic chains}
\medskip

We now consider a kinetic mechanism  in accordance with the assumptions  underlying the Lotka-Volterra prey-predator equations. We 
represent resources by $A$, prey by $N_1$ and predators by $N_2$.
In order to be able to derive a conservation law we introduce a recycling mechanism depending upon two unknown parameters, $\beta_1$ and $\beta_2$.
Under these conditions, the prey-predator mechanism is:
$$\eqalign{
A+N_1 \mathrel{\mathop \rightarrow^{r_1}} (1+e_1) N_1 \qquad &N_1 \mathrel{\mathop \rightarrow^{d_1}} \beta_1A \cr
N_1+N_2 \mathrel{\mathop \rightarrow^{r_2}} (1+e_2) N_2 \qquad &N_2 \mathrel{\mathop \rightarrow^{d_2}} \beta_2A \cr
}\eqno(5.1)
$$
where $e_1$ and $e_2$ are conversion factors, and $d_1$ and $d_2$ are death rates.

The dynamic equations  for mechanism (5.1) become,
$$\eqalign{
{dA \over dt}&=-r_1 A N_1+d_1 \beta_1N_1 +d_2 \beta_2N_2 \cr
{dN_1 \over dt}&=e_1 r_1 A N_1-r_2 N_1 N_2-d_1 N_1 \cr
{dN_2 \over dt}&=r_2 e_2 N_1 N_2-d_2 N_2 \cr
}\eqno(5.2)
$$

Imposing the conservation law,
$$
A+\gamma_1  N_1 +\gamma_2  N_2  =K\eqno(5.3a)
$$
and after derivation, by (5.2), we obtain 
$$
\beta_1=\gamma_1=1/e_1\, , \beta_2=\gamma_2=1/(e_1 e_2)\eqno(5.3b)
$$

Using the conservation law (5.3) to eliminate $A$ from (5.2),  
the equations for the prey-predator mechanism become
$$\eqalign{
{dN_1 \over dt}&=e_1 r_1\left(K-{N_1 \over e_1}-{N_2 \over e_1 e_2}\right)N_1-r_2 N_1 N_2-d_1 N_1 \cr
{dN_2 \over dt}&=r_2 e_2 N_1 N_2-d_2 N_2 \cr
}\eqno(5.4)
$$

These equations are functionally equivalent to the usual Lotka-Volterra equations. However, due to mass conservation, the parameters no longer have exactly the same meaning. In fact, prey growth is not only controlled by predation but also by the fact that predators retain nutrient within them.

We can generalize the preceding mechanism to trophic chains of arbitrary length:

$$\eqalign{
A+N_1 \mathrel{\mathop \rightarrow^{r_1}} (1+e_1) N_1 \qquad & N_1 \mathrel{\mathop \rightarrow^{d_1}} {1 \over e_1}A \cr
N_1+N_2 \mathrel{\mathop \rightarrow^{r_2}} (1+e_2)N_2 \qquad &N_2 \mathrel{\mathop \rightarrow^{d_2}} {1 \over {e_1 e_2}}A \cr
\vdots \cr
N_{n-1}+N_n \mathrel{\mathop \rightarrow^{r_n}} (1+e_n)N_n \qquad & N_n \mathrel{\mathop \rightarrow^{d_n}} {1 \over {e_1 ... e_n}}A \cr
}\eqno(5.5)
$$

This system has the conservation law:
$$
A+{N_1 \over e_1}+{N_2 \over e_1 e_2}+...+{N_n \over e_1 e_2 ... e_n} =K\eqno(5.6)
$$

Using the conservation law to eliminate $A$, the dynamical equations become
$$\eqalign{
{dN_1 \over dt}&=e_1 r_1\left(K-{N_1 \over e_1}-{N_2 \over e_1 e_2}-...{N_n \over e_1 e_2...e_n}\right)N_1-r_2 N_1 N_2-d_1 N_1 
\cr
{dN_i \over dt}&=r_i e_i N_{i-1} N_i-d_i N_i \, , i=2,\ldots , n\cr}\eqno(5.7)
$$

For trophic chains of length greater than two, the equation for the basal species is functionally different from the Lotka-Volterra food chain. The basal species is controlled by all the other species, since they are all retaining nutrient (this effect increases with the increase in nutrient retained in the nonbasal trophic levels). 

\bigskip
{\bf 6. Trophic chains with internal states}
\medskip

We now derive the basic population dynamics equations for the time evolution 
of $n$ species in a food chain, assuming that all  the species involved have some refractory time, during which they are not able to consume resources.  

Consider a food chain with $n$ species,
$N_1,\ldots , N_n$ and a primary resource  $A$. Suppose in addition that
each species has two states $N_{is}$ and $N_{ih}$, where subscripts $s$ and 
$h$ stand, respectively, for "searching for prey" and "handling for prey".
Introducing this distinction we have $N_i=N_{is}+N_{ih}$. With a 
characteristic time $t_{ih}$, handling predators finish handling their prey and return to the searching state. 
Reproduction transforms searching into handling predactors. Under these conditions, the mechanism for the trophic chain is
$$\eqalign{
A+N_{1s} \mathrel{\mathop \rightarrow^{r_1}} (1+e_1) N_{1h} \qquad & N_{1s} \mathrel{\mathop \rightarrow^{d_1}} \beta_1A \cr
N_{1h} \mathrel{\mathop \rightarrow^{t_{1h}^{-1}}} N_{1s} \qquad & N_{1h} \mathrel{\mathop \rightarrow^{d_1}} \beta_1 A \cr
N_{1s}+N_{2s} \mathrel{\mathop \rightarrow^{r_2}} (1+e_2)N_{2h} \qquad &N_{2h} \mathrel{\mathop \rightarrow^{d_2}} \beta_2A \cr
N_{1h}+N_{2s} \mathrel{\mathop \rightarrow^{r_2}} (1+e_2)N_{2h} \qquad & \cr
N_{2h} \mathrel{\mathop \rightarrow^{t_{2h}^{-1}}} N_{2s} \qquad & N_{2s} \mathrel{\mathop \rightarrow^{d_2}} \beta_2 A \cr
\vdots &\cr
N_{n-1s}+N_{ns} \mathrel{\mathop \rightarrow^{r_n}} (1+e_n)N_{nh} \qquad & N_{nh} \mathrel{\mathop \rightarrow^{d_n}} \beta_nA \cr
N_{n-1h}+N_{ns} \mathrel{\mathop \rightarrow^{r_n}} (1+e_n)N_{nh} \qquad & \cr
N_{nh} \mathrel{\mathop \rightarrow^{t_{nh}^{-1}}} N_{ns} \qquad & N_{ns} \mathrel{\mathop \rightarrow^{d_n}} \beta_n A
}\eqno(6.1)
$$
Applying the mass action law  of \S 2, we obtain the evolution equations,
$$\eqalign{
{dA\over dt}&=-r_1AN_{1s}+\sum_{i=1}^{n}\beta_i d_i(N_{is}+N_{ih}) \cr
{dN_{1s}\over dt}&=-r_1AN_{1s}+t_{1h}^{-1}N_{1h}-d_1N_{1s}-r_2 N_{1s}N_{2s}\cr
{dN_{1h}\over dt}&=r_1(1+e_1)AN_{1s}-t_{1h}^{-1}N_{1h}-d_1N_{1h}-r_2 N_{1h}N_{2s}\cr
 \vdots &\cr
{dN_{is}\over dt}&=-r_i(N_{i-1s}+N_{i-1h})N_{is}+t_{ih}^{-1}N_{ih}-d_iN_{is}-r_{i+1} N_{is}N_{i+1s}\cr
{dN_{ih}\over dt}&=r_i(1+e_i)(N_{i-1s}+N_{i-1h})N_{is}-t_{ih}^{-1}N_{ih}-d_iN_{ih}-r_{i+1} N_{ih}N_{i+1s}\cr
 \vdots &\cr
{dN_{ns}\over dt}&=-r_n(N_{n-1s}+N_{n-1h})N_{ns}+t_{nh}^{-1}N_{nh}-d_nN_{ns}\cr
{dN_{nh}\over dt}&=r_n(1+e_n)(N_{n-1s}+N_{n-1h})N_{ns}-t_{nh}^{-1}N_{nh}-d_nN_{nh}\cr 
}\eqno(6.2)
$$
where, $i=2,\ldots , n-1$. We now impose a conservation law of the form,
$$
{d A\over dt}+ \gamma_1 \left({dN_{1s}\over dt}+{dN_{1h}\over dt}\right)+\cdots +\gamma_n \left({dN_{ns}\over dt}+{dN_{nh}\over dt}\right)=0\eqno(6.3)
$$
where $\gamma_i$ are unknown parameters. Introducing (6.2) into (6.3), and
solving for $\gamma_i$ and $\beta_i$, we obtain
$$
\gamma_i={1\over e_1\ldots e_i}\, , \, \beta_i=\gamma_i\, ,\, i=1,\ldots , n\eqno(6.4)
$$
Hence, the conservation law is
$$
A+{1\over e_1}N_1+\ldots +{1\over e_1\ldots e_n}N_n=K\eqno(6.5)
$$
where $K$ is the carrying capacity, and $N_i=N_{is}+N_{ih}$.

Let us now introduce a steady state assumption over the behavioral states 
$N_{ih}$:
$$
{dN_{ih}\over dt}=0\, , i=1, \ldots ,n\eqno(6.6)
$$
With, $N_i=N_{is}+N_{ih}$, and writing the system of equations (6.2) as
a function of $N_n$ and $N_{ns}$, we obtain the dynamic equations for  species densities,
$$\eqalign{
{dN_{1}\over dt}&=e_1r_1(K-{1\over e_1}N_1-\ldots -{1\over e_1\ldots e_n}N_n)N_{1s}-r_2 N_{1}N_{2s}-d_1N_{1}\cr
 \vdots &\cr
{dN_{i}\over dt}&=e_ir_iN_{is}N_{i-1}-r_{i+1} N_{i}N_{i+1s}-d_iN_{i}\cr
 \vdots &\cr
{dN_{n}\over dt}&=e_nr_nN_{ns}N_{n-1}-d_nN_{n}\cr
}\eqno(6.7)
$$
In order to eliminate $N_{is}$ from (6.7), we solve the system of equations
(6.6) together with the relation  $N_i=N_{is}+N_{ih}$. Therefore, (6.7) together
with (6.6) are the general equations for a trophic chain. 

Let us now analyze the case $n=2$. In this case, (6.7) and (6.6) simplify to
$$
\eqalign{
{dN_{1}\over dt}&=e_1r_1(K-{1\over e_1}N_1-{1\over e_1e_2}N_2)N_{1s}-r_2 N_{1}N_{2s}-d_1N_{1}\cr
{dN_{2}\over dt}&=e_2r_2N_{1}N_{2s}-d_2N_{2}\cr
{dN_{1h}\over dt}&=r_1(1+e_1)AN_{1s}-t_{1h}^{-1}(N_1-N_{1s})-d_1(N_1-N_{1s})-r_2 (N_1-N_{1s})N_{2s}=0\cr
{dN_{2h}\over dt}&=r_2(1+e_2)N_{1}N_{2s}-t_{2h}^{-1}(N_2-N_{2s})-d_2(N_2-N_{2s})=0\cr
}\eqno(6.8)
$$
Solving the two last equations in (6.8) for $N_{1s}$ and $N_{2s}$, we obtain,
$$
\eqalign{
N_{2s}&=N_2{t_{2h}^{-1}+d_2\over t_{2h}^{-1}+d_2+r_2(1+e_2) N_1}\cr
N_{1s}&=N_1{t_{1h}^{-1}+d_1+r_2N_2{t_{2h}^{-1}+d_2\over t_{2h}^{-1}+d_2+r_2(1+e_2) N_1}\over t_{1h}^{-1}+d_1+r_1(1+e_1) (K-{1\over e_1}N_1-{1\over e_1e_2}N_2)+r_2N_2{t_{2h}^{-1}+d_2\over t_{2h}^{-1}+d_2+r_2(1+e_2) N_1}}\cr
}\eqno(6.9)
$$
Therefore, system (6.8) simplifies to
$$
\eqalign{
{dN_{1}\over dt}&=e_1r_1(K-{1\over e_1}N_1-{1\over e_1e_2}N_2)N_{1s}-r_2 N_{1}N_{2s}-d_1N_{1}\cr
{dN_{2}\over dt}&=e_2r_2N_{2s}N_{1}-d_2N_{2}\cr
}\eqno(6.10)
$$
where $N_{1s}$ and $N_{2s}$ are given by (6.9). 

Equation (6.10) describes 
a trophic chain with a prey, a predator  and a renewable resource. The carrying
capacity of the system is the constant $K$. The introduction of the intermediate states $N_{ih}$, specifying the existence of a refractory time where species are not consuming resources, implies a functional response of Holling type II, with ecologically meaningful parameters. If we consider that the basal species in 
$N_1$ has negligible handling time, $t_{1h}^{-1}\to \infty$, then, by (6.9),
$N_{1s}\to N_{1}$, and (6.10) reduces to
$$
\eqalign{
{dN_{1}\over dt}&=e_1r_1(K-{1\over e_1}N_1-{1\over e_1e_2}N_2)N_{1}-r_2 N_{1}N_2{t_{2h}^{-1}+d_2\over t_{2h}^{-1}+d_2+r_2(1+e_2) N_1}-d_1N_{1}\cr
{dN_{2}\over dt}&=e_2r_2N_{1}N_2{t_{2h}^{-1}+d_2\over t_{2h}^{-1}+d_2+r_2(1+e_2) N_1} -d_2N_{2}\cr
}\eqno(6.11)
$$
wnich is a modified Rosenzweig-MacArthur (1963) predation model. 

In the case  the handling times $t_{1h}^{-1}$ and $t_{2h}^{-1}$ go to
infinity, by (6.9), $N_{is}$ converges to $N_{i}$, and (6.10) reduces to
the modified Lotka-Volterra equation (5.4).

We now compare the time evolution of system (6.10) with the modified 
prey-predator Lotka-Volterra system (5.4). We take the parameter values:
$r_1=1.0$, $r_2=0.2$, $e_1=1$, $e_2=1$, $d_1=0.1$, $d_2=0.2$, $t_{1h}^{-1}=1.0$, $t_{2h}^{-1}=2.0$,
$K=8$, $N_1(0)=1.0$ and $N_2(0)=0.1$.  In Fig. 3, we depict the time evolution of the two models, for prey and predators.
In the case of model equations (5.4), the system attains an equilibrium
value given by $N_1=d_2/e_2 r_2$ and $N_2=(r_2(K e_1 r_1-d_1)-d_2 r_1)/(r_2(r_1+e_2 r_2))$. However, for system (6.10), we obtain stable oscillations
corresponding to a limit cycle in phase space.

For the general case of an arbitrary trophic chain we have equation (6.2) together with the carrying capacity relation (6.5), and, eventually,
the steady state conditions (6.6).

\bigskip
{\bf 7. Conclusions}
\medskip

We have developed a systematic formalism, based on chemical kinetics, for the derivation of equations in population dynamics based on the mechanisms of interaction between individuals. We have started from the simplest equation, the logistic, and then introduced successive levels in a trophic chain. The dynamical equations are derived using the laws of mass action and mass conservation, and, when necessary, a steady state assumption. Our approach includes any kind of trophic interaction between species and resources and internal states.  This approach   gives a consistent mechanistic basis for the
derivation of the trophic chain  equations of population biology, making it possible to settle several controversies in Ecology. Moreover, this formalism  allows the precise development of more complicated models, with the introduction of more mechanisms and interactions, allowing the development of extensions of the logistic equation with precise applicability conditions.

This approach has the advantage that all the parameters have an {\it a priori} biological meaning. Moreover, we show the necessity of the introduction of a conservation law relating populations and primary resource densities. In this context, the carrying capacity parameter is the value assumed by the conservation law, and appears in all population dynamics models, from one species to arbitrary food webs. However, the numerical equilibrium value attained by the populations only equals the carrying capacity in the case of the logistic equation.

The use of the mass action law, together with its microscopic foundation, allows the development of models with stochastic fluctuation around a mean value (Bailey, 1964; Haken, 1983), as well as the development of equations with different growth forms and the introduction of individual variability.

Finally, one of the consequences of the systematic use of the mass action law is the proof of the empirical observation that the growth rate $r$, and the carrying capacity $K$ in the logistic equation, both increase proportionally to enrichment.

\vfill \eject
\bigskip
\centerline{\bf References}

\noindent Allee, W. C., Emerson, A. E., Park, O., Park, T. and Schmidt, K. P., 1949. Principles of Animal Ecology. Saunders, Philadelphia, Pennsylvania.
\medskip

\noindent Anderson, R. M. and May, R. M., 1991. Infectious Diseases of Humans: Dynamics and Control. Oxford University Press, Oxford.
\medskip

\noindent Arditi, R. and Ginzburg, L. R., 1989. Coupling in predator-prey dynamics: ratio-dependence. J. Theor. Biol., 139: 311-326.
\medskip

\noindent Bailey, N. T. J., 1964. The Elements of Stochastic Processes: With 
Applications to the Natural Sciences. J. Wiley and Sons, New York.
\medskip

\noindent Banks, R. B., 1994. Growth and Diffusion Phenomena: Mathematical Frameworks and Applications. Texts in Applied Mathematics Vol. 14. Springer-Verlag, Berlin, 19+455 pp.
\medskip

\noindent Berryman, A. A., 1992. The origins and evolution of predator-prey theory. Ecology, 73: 1530-1535.
\medskip

\noindent Berryman, A. A. and Millstein, J. A., 1990. Population Analysis System: POPSYS Series 1, One-Species Analysis (Version 2.5). Ecological Systems Analysis, Pullman, Washington. 
\medskip

\noindent Berryman, A. A., Guttierrez, A. P. and Arditi, R., 1995a. Credible, parsimonious and useful predator-prey models: a reply to Abrams, Gleeson, and Sarnelle. Ecology, 76: 1980-1985.
\medskip

\noindent Berryman, A. A., Michalski, J., Gutierrez, A. P. and Arditi, R., 1995b. Logistic theory of food web dynamics. Ecology, 76: 336-343.
\medskip

\noindent Borghans, J. A. M., De Boer, R. J. and Segel, L. A., 1996. Extending the quasi-steady state approximation by changing variables, Bull. Math. Biol., 58: 43-63.
\medskip

\noindent Clark, C. W., 1990. Mathematical Bioeconomics. John Wiley and Sons, New York, New York.
\medskip

\noindent DeAngelis, D. L., Bartell, S. M. and Brenkert, A. L., 1989. Effects of nutrient recycling and food-chain length on resilience. Am. Nat., 134: 778-805. 
\medskip

\noindent Gause, G. F., 1934. The Struggle for Existence. Williams and Wilkins, New York, New York.
\medskip

\noindent Getz, W. M., 1984. Population dynamics: a per capita resource approach. J. Theor. Biol., 108: 623-643.
\medskip

\noindent Haken, H., 1983. Synergetics  (3rd edition). Springer-Verlag, Berlin.
\medskip

\noindent Holling, C. S., 1959. The components of predation as revealed by a study of small mammal predation of the European pine sawfly. The Canadian Entomologist, 91: 293-320.
\medskip

\noindent  Kooi, B. W., Boer, M. P. and Kooijman, S. A. L. M., 1998. On the use of the logistic equation in models of food chains. Bull. Math. Biol., 60: 231-246.
\medskip

\noindent Maurer, B. A., 1998. Ecological Science and Statistical Paradigms: 
At the Threshold. Science, 279: 502-503. 
\medskip

\noindent May, R. M., 1974. Stability and Complexity in Model Ecosystems (2nd Ed.). Monographs in Population Biology Vol. 6. Princeton University Press, Princeton, New Jersey.
\medskip

\noindent Metz, J. A. J. and de Roos, A. M., 1992. The role of physiologically structured population models within a general individual-based modelling perspective. In: D. L. DeAngelis and L. J. Gross (Editors), Individual-Based Models and Approaches in Ecology: Populations, Individuals and Ecosystems, Chapman and Hall, New York, New York, pp. 88-111. 
\medskip

\noindent Metz, J. A. J. and Diekmann, O. (Editors), 1986. The Dynamics of Physiologically Structured Populations. Lecture Notes in Biomathematics 68. Springer-Verlag, Berlin.  
\medskip

\noindent Michalski, J., Poggiale, J.-Ch., Arditi, R. and Auger, P. M., 1997. Macroscopic dynamic effects of migrations in patchy predator-prey systems. J. Theor. Biol., 185: 459-474.
\medskip

\noindent Nicolis, G. and Prigogine, I., 1997. Self-Organization in Nonequilibrium Systems. J. Wiley and Sons, New York.
\medskip

\noindent Oksanen, L., Fretwell, S. D., Arruda, J. and Niemela, P., 1981. Exploitation ecosystems in gradients of primary productivity. Am. Nat., 140: 938-960.
\medskip

\noindent O'Neill, R. V., DeAngelis, D. L., Waide, J. B. and Allen, T. F. H., 1986. A Hierarchical Concept of Ecosystems. Monographs in Population Biology Vol. 23. Princeton University Press, Princeton, New Jersey, 7+253 pp.
\medskip

\noindent Rosenzweig, M. and MacArthur, R. H. 1963. Graphical representation and stability conditions of predator-prey interactions. Am. Nat., 107: 275-294.
\medskip

\noindent Royama, T., 1971. A comparative study of models for predation and parasitism. Res. Popul. Ecol. Supp. 1.
\medskip

\noindent Schlegel, H. G., 1992. General Microbiology (7th edition). Cambridge
University Press, Cambridge.
\medskip

\noindent Segel, L. A., 1988. On the validity of the steady state assumption of enzyme kinetics. Bull. Math. Biol., 6: 579-593.
\medskip

\noindent Segel, L. A. and Slemrod, M., 1989. The quasi steady state assumption: a case study in perturbation. SIAM Rev., 31: 466-477.
\medskip

\noindent Solow, A. R., 1995. Fitting population models to time series data. In: T. M. Powell and J. H. Steele (Editors), Ecological Time Series, Chapman and Hall, New York, New York, pp. 20-27.
\medskip

\noindent Stiefenhofer, M. 1998. Quasi-steady-state approximation for chemical reaction networks. J. Math. Biol. 36, 593-609.
\medskip

\noindent Thomas, W. R., Pomerantz, M. J. and Gilpin, M. E., 1980. Chaos, assymetric growth and group selection for dynamical stability. Ecology, 61: 1312-1320.
\medskip

\noindent Van Kampen, N. G., 1992. Stochastic Processes in Physics and Chemistry, Elsevier, Amsterdam, 1992.
\medskip

\vfill\eject
\centerline{\bf Figure Captions}

\noindent {\bf Figure 1:} Solution $N(t)$  of the logistic type models
(3.3), (3.6) and (3.8). The  parameter values are:
$e=1$, $r_0=1$, $d=0.2$,
$K=20$, $N(0)=0.1$ and $A(0)=(K-N(0))/e=19.9$. 
\bigskip

\noindent {\bf Figure 2:} Solution $N(t)$  for the logistic and Monod type model 
(4.8). The  parameter values are:
$e=1$, $r_1=1$, $r_2=10$, $d=0.2$,
$K=20$, $N(0)=0.1$, $A(0)=(K-N(0))/e=19.9$ and $\delta=0.196$. In the limit
$\delta \to 0$ (or $r_2\to \infty$) the solution of the Monod type
model approches the solution of the logistic equation.
\bigskip

\noindent {\bf Figure 3:} Solutions $N(t)$  of the modified Lotka-Volterra 
model (5.4) and  trophic model (6.10). The  parameter values are:
$r_1=1.0$, $r_2=0.2$, $e_1=1$, $e_2=1$, $d_1=0.1$, $d_2=0.2$, $t_{1h}^{-1}=1.0$, $t_{2h}^{-1}=2.0$,
$K=8$, $N_1(0)=1.0$ and $N_2(0)=0.1$.

\bye